\documentclass[aps,twocolumn,showpacs,prl,twoside,amssymb,amsmath]{revtex4}
\usepackage{amssymb}
\usepackage{graphicx}
\usepackage{amsmath}
\usepackage{colordvi}
\usepackage{bbm}
\usepackage{amsmath}

\newcommand{\tr}{{\rm Tr}}

\newcommand{\mU}{{\mathcal U}}

\begin{document}

\title{Witnessing the quantum discord of all the unknown states}

\author{Sixia Yu$^{1,2}$}
\email{cqtys@nus.edu.sg}
\author{Chengjie Zhang$^{1}$}
\author{Qing Chen$^{1,2}$}
\author{C.H. Oh$^{1}$}
\email{phyohch@nus.edu.sg}
\affiliation{$^1$Centre for quantum technologies, National University of Singapore, 3 Science Drive 2, Singapore 117543\\
$^2$Hefei National Laboratory for Physical Sciences at Microscale and Department of Modern Physics, \\ University of Science and Technology of China, Hefei, Anhui 230026, China}

\begin{abstract}
 Like entanglement, quantum discord quantifies the quantum correlations. Unlike entanglement, whose detection is extremely difficult, the quantum discord of an arbitrary bipartite state allows itself to be detected perfectly by a single observable, the quantum discord witness. In particular, we report a single observable whose  expectation value  provides a necessary and sufficient condition for the vanishing quantum discord of an arbitrary bipartite unknown state with four copies. A quantum circuit is designed to measure the quantum discord witness by using only local qubit-measurements.
\end{abstract}


\maketitle
Entanglement is an essential resource in almost all quantum computing and informational processing tasks. In certain quantum computing tasks, however, the quantum advantages can be gained without entanglement. One typical example is the deterministic quantum computation with one qubit \cite{DQC1} in which the {\it quantum discord}, introduced by Ollivier and Zurek \cite{discord1} and independently by Henderson and Vedral \cite{discord2}, is proposed to be responsible for the quantum speedup \cite{Datta}. Ever since, the quantum discord has found numerous applications in, e.g.,  characterizing the complete positivity of a map \cite{CP}, local broadcasting of the quantum correlations \cite{broadcast,luo}, and indicating the phase transitions \cite{phase}. The operational interpretations of the quantum discord proposed by Cavalcanti \textit{et al.} \cite{merg1} and by V. Madhok and A. Datta \cite{merg2} have established firmly its status as another essential resource.

As a nonnegative number quantifying the quantum correlations beyond entanglement, the quantum  discord of a bipartite state $\varrho_{AB}$ is defined by \cite{discord1,discord2}
\begin{equation}\label{def}
D_A(\varrho_{AB}):=\min_{\{E_i^A\}}\sum_{i}p_iS(\varrho_{B|i})+S(\varrho_A)-S(\varrho_{AB}),
\end{equation}
where $S(\varrho)=-\mathrm{Tr}(\varrho\log_2\varrho)$ denotes the von Neumann entropy and the minimum is taken over all the positive operator valued measures (POVMs) $\{E_i^A\}$ on the subsystem $A$ with $p_i=\tr (E_i^A\varrho_{AB})$ being the probability of the $i$-th outcome and $\varrho_{B|i}=\tr_A(E_i^A\varrho_{AB})/p_i$ being the conditional state of subsystem $B$. The minimum can also be taken over all the von Neumann measurements~\cite{discord1}. These two definitions produce in general different values but they are identical for zero-discord states: $\mathcal{D}_A(\varrho_{AB})=0$ if and only if there exist a complete and orthonormal basis $\{|k\rangle_A\}$ for subsystem $A$ and a set of density matrices $\varrho^k_B$ of subsystem $B$ such that
\begin{equation}\label{zs}
\varrho_{AB}=\sum_k p_k |k\rangle\langle k|_A\otimes\varrho_{k}^{B}.
\end{equation}

One of the fundamental issue is therefore how to detect those states with vanishing quantum discord, either known or unknown. For a known bipartite state many criteria for non-zero quantum discord have been proposed. For examples Ferraro {\it et al.} \cite{almost} proposed the commutativity of the reduced density matrix to be a necessary condition. Bylicka and Chru\'{s}ci\'{n}ski  \cite{2N} proposed another necessary condition for $2\times N$ systems in terms of strong positive partial transpose. Specially Rahimi and SaiToh  \cite{Rahimi} introduced an example of nonlinear witness for the nonzero quantum discord. Recently a necessary and sufficient condition is proposed by  Daki\'{c} {\it et al.} \cite{condition1} and Chen  {\it et al.} \cite{chen}  in the form of local commutativity, which solves completely the problem of detecting the quantum discord in the case of known states.  In the case of unknown state Zhang {\it et al.} \cite{zhang} constructed a single observable whose vanishing expectation value provides a sufficient condition, which becomes also necessary for $2\times N$ states, for the vanishing discord.

In this Letter we solve completely the problem of witnessing the  quantum discord of an arbitrary bipartite unknown state. In particular we propose a single observable, referred to as the quantum discord witness, to detect the quantum discord of an unknown state, provided that there are four copies of the state. The vanishing expectation value of the quantum discord witness provides a necessary and sufficient condition for a vanishing quantum discord. Moreover we have designed a quantum circuit to measure the discord witness by using only local qubit measurements.

We consider a general bipartite state $\varrho_{AB}$ in a $d_A\times d_B$ system. To detect its quantum discord we note that one crucial lesson learned from previous results \cite{Rahimi, zhang} is that nonlinear instead of linear witnesses (with respect to the density matrix) should be considered. Also we note that the expectation values of an observable on multi copies of a state is naturally nonlinear. Furthermore the quantum discord is invariant under local unitary (LU) transformations and therefore the observable witnessing the quantum discord should also be LU-invariant.

For a given bipartite state $\varrho_{AB}$, a polynomial LU invariant of degree $k$  is given by $\mathrm{Tr}(U^AU^B\varrho^{\otimes k}_{AB})$, where $U^{A(B)}$ is some permutation operator acting on $k$ copies of subsystem $A(B)$ \cite{invariants}. In what follows we shall consider only $k=4$ copies of the state and label them with numbers from 1 to 4. As examples the permutation operator $U^B$ may be taken as $V^B_{12} V_{34}^B$ or $V^B_{13} V_{24}^B$,  where
\begin{equation}
V_{ij}^B=\sum_{n_1,n_2=0}^{d_B-1}|n_1,n_2\rangle\langle n_2,n_1|_{ij}=\sum_{\mu=0}^{d_B^2-1}G_\mu^{B_i}\otimes G_\mu^{B_j}
\end{equation}
is the swapping operator acting on two copies of qudit $B$ labeled with  $i,j=1,2,3,4$. Here we have introduced  a complete set of local orthogonal observables \cite{LOO} $\{G_\mu^{B}\mid \mu=0,1,2,\ldots d^2_{B}-1\}$ satisfying $\mathrm{Tr}(G_\mu^{B} G_\nu^{B})=\delta_{\mu\nu}$ for qudit $B$. The permutation operator $U^A$ may be the cyclic permutation operator
\begin{equation}
X_A=\sum_{n_1,n_2,n_3,n_4=0}^{d_A-1}|n_1,n_2,n_3,n_4\rangle\langle n_2,n_3,n_4,n_1|
\end{equation}
acting on four copies of qudit $A$. It is obvious that $X_A=V_{12}^AV_{23}^AV_{34}^A$. Our main result is:

{\it Theorem } A $d_A\times d_B$ bipartite state $\varrho_{AB}$ has a vanishing quantum discord, i.e, $D_A(\varrho_{AB})=0$, if and only if $\tr (W\varrho_{AB}^{\otimes 4})=0$ where
\begin{equation}
W=\frac12(X_A+X_A^\dagger)(V_{13}^BV_{24}^B-V_{12}^BV_{34}^B).
\end{equation}

{\it Proof } The density matrix $\varrho_{AB}$ has a partial expansion $\varrho_{AB}=\sum_{\mu}\varrho^A_\mu\otimes G_\mu^B$ with $\varrho^A_\mu=\tr_B(\varrho_{AB}G_\mu^B)$ being Hermitian. For any four operators $\varrho_i$ $(i=1,2,3,4)$ acting on qudit $A$ it holds $\tr (X_A(\varrho_1\otimes\varrho_2\otimes\varrho_3\otimes\varrho_4))=\tr(\varrho_4\varrho_3\varrho_2\varrho_1)$.
Straightforward calculations yield
\begin{eqnarray}
\tr(W\varrho_{AB}^{\otimes 4})&=&\sum_{\mu,\nu=0}^{d_B^2-1}\tr\left(\varrho^A_\mu\varrho^A_\nu\right)^2-\tr
\left((\varrho^A_\mu)^2(\varrho^A_\nu)^2\right)\nonumber\\
&=&-\frac12\sum_{\mu,\nu=0}^{d_B^2-1}\tr\left(i[\varrho^A_\mu,\varrho^A_\nu]\right)^2.\label{mi}
\end{eqnarray}
If $D_A(\varrho_{AB})=0$ then $\varrho_{AB}$ is of the form in Eq.(\ref{zs}) and  $[\varrho^A_\mu,\varrho^A_\nu]=0$ for all $\mu,\nu$ leading to $\tr(W\varrho_{AB}^{\otimes 4})=0$. If $\tr(W\varrho_{AB}^{\otimes 4})=0$ then $[\varrho^A_\mu,\varrho^A_\nu]=0$ for all $\mu,\nu$ since each of the terms in the sum over $\mu,\nu$ in Eq.(\ref{mi}) is nonnegative. Therefore commuting operators $\varrho_\mu^A$ have a common set of eigenstates so that the state $\varrho_{AB}$ can be brought into the form in Eq.(\ref{zs}), i.e., $D_A(\varrho_{AB})=0$.\hfill Q.E.D.

\begin{figure}
\begin{center}
\includegraphics[scale=0.6]{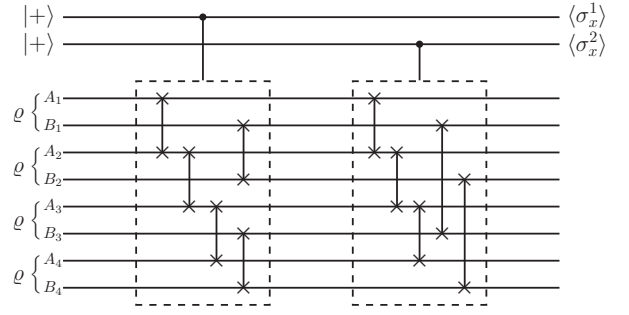}
\caption{Quantum circuit to measure the quantum discord witness.
Two auxiliary qubits, initially prepared in state $|+\rangle$, are the sources of a series of controlled-swapping gates, which are represented by connected crosses. All the swapping gates  in the dashed-boxes on the left and right are controlled by qubits 1 and 2 respectively. Two local $\sigma_x$-measurements constitute a measurement of the quantum discord witness. }\label{1}
\end{center}
\end{figure}

In order to measure the expectation value of the discord witness $W$ we have designed a quantum circuit which is illustrated in Fig.1. There we have introduced two ancilla qubits that are initially prepared in the the +1 eigenstate
$[+]=|+\rangle\langle +|$ of $\sigma_x$. That is to say the initial state of the whole system is $[+]_1\otimes[+]_2\otimes\varrho_{AB}^{\otimes 4}$. After applying a series of controlled swapping gates to the whole system  with two qubits as sources, which are given by $\mU_a=[0]_a\otimes I+[1]_a\otimes U_a$ with $a=1,2$ and $U_1= X_AV^B_{12}V^B_{34}$ and $U_2=X_AV^B_{13}V^B_{24}$ as shown in Fig.1, the reduced density matrix of  qubit $a$ is $(1+\sigma_x^a\tr(\frac12(U_a+U_a^\dagger)\varrho^{\otimes 4}_{AB}))/2$. Therefore if we make two local $\sigma_x$-measurements then the resulting expectation values $\langle \sigma_{x}^a\rangle$ determine the expectation value of the discord witness according to
\begin{equation}\label{meas}
\tr(W\varrho_{AB}^{\otimes 4})=\langle \sigma_{x}^2\rangle-\langle \sigma_{x}^1\rangle.
\end{equation}

In the case of $d_A=2$ the expectation value $\tr (W\varrho_{AB}^{\otimes 4}) $ coincides with the expectation value
$\tr (W_{2\times N}\varrho_{AB}^{\otimes 4})$ of the discord witness $W_{2\times N}$ for $2\times N$ system introduced in \cite{zhang}, as expected. We note that the condition for zero discord is an equality, i.e., $\tr (W\varrho_{AB}^{\otimes 4})=0$, which means that at least one parameter of the zero-discord state $\varrho_{AB}$ must not be free. Therefore the states with zero-discord are of zero measure  as shown in \cite{almost}.

We acknowledge the financial support of A*STAR Grant WBS No. R-144-000-189-305, and the financial support from CQT project WBS: R-710-000-008-271.

\end{document}